\documentclass{PoS}
\usepackage{url}
\newcommand{\mnras}{Monthly Notices of the Royal Astronomical Society}
\newcommand{\apj}{The Astrophysical Journal}
\newcommand{\apjl}{The Astrophysical Journal Letters}

\newcommand{\aap}{Astronomy and Astrophysics}

\newcommand{\mwl}{multi-wavelength}
\newcommand{\pks}{PKS~2155--304}
\newcommand{\Bf}{\textit{B}}
\newcommand{\Vf}{\textit{V}}
\newcommand{\Rf}{\textit{R}}

\newcommand{\hess}[1]{H.E.S.S. #1}
\newcommand{\gae}{\lower 2pt \hbox{$\, \buildrel {\scriptstyle >}\over {\scriptstyle
\sim}\,$}}
\newcommand{\lae}{\lower 2pt \hbox{$\, \buildrel {\scriptstyle <}\over {\scriptstyle
\sim}\,$}}
\newcommand{\xray}{\textit{X}--ray}
\newcommand{\gray}{$\gamma$--ray}
\newcommand{\dgr}{$^\circ$}

\title{The Optical Polarization Variability of PKS~2155--304: Correlations at Low and High Energies}

\ShortTitle{The Optical Polarization Variability of PKS~2155--304}

\author{\speaker{N.~W. Pekeur}\thanks{The authors would like to extend their thanks to Philip Edwards and Jamie Stevens for access to radio observations of \pks{} from the Australia Telescope Compact Array (ATCA) Calibrator Survey.}\\
        Department of Astronomy, University of Cape Town, Rondebosch 7701, SA\\
        E-mail: \email{nikki.pekeur@ast.uct.ac.za}}

\author{R. Taylor \\
			   Department of Astronomy, University of Cape Town, Rondebosch 7701, SA;\\
               Department of Physics and Astronomy, University of the Western Cape, Belville 7530, SA\\
        E-mail: \email{russ@ast.uct.ac.za}}

\author{R.~C. Kraan-Korteweg \\
        Department of Astronomy, University of Cape Town, Rondebosch 7701, SA\\
        E-mail: \email{kraan@ast.uct.ac.za}}

\abstract{Gamma-ray emitting blazars have been the targets of many recent or ongoing monitoring programmes. The resulting \mwl{} data archives are ideal for studying the quiescent and variable behaviour of the blazar \pks. Here, observations of the optical polarization of the source between 2009 and 2014 ($> 5$ years) are presented, together with overlapping radio, photometric and gamma-ray measurements. During this time the source displayed significant variability at all wavelengths, with several prominent flares seen for the optical and gamma-ray intensity. These flares are typically associated with large rotations ($\gae 80^\circ$) in the polarization angle, during which the polarization angle appears to be oriented parallel to the position angle of the spatially resolved radio jet. Our aim is to determine if these large polarization angle rotations are correlated to the \mwl{} outbursts seen in \pks{} in order to examine how changes in the polarization angle (and its orientation relative to the jet) relate to changes in the spectral energy distribution of the source.}
\FullConference{4th Annual Conference on High Energy Astrophysics in Southern Africa\\
		25-27 August, 2016\\
		Cape Town, South Africa}

\begin{document}

\section{Introduction}
The active galaxy \pks{} is an archetypal blazar, for which the relativistic jet is along the line-of-sight of the observer and the jet has a small opening angle, that is associated with a compact, flat-spectrum radio source. Located at a redshift $z=0.116$, it is highly luminous, with an apparent \Vf--band magnitude $m_V =13^\mathrm{mag}$ \cite{Falomo1993}. At low energies (radio to soft \xray), relativistic charged particle acceleration in the magnetic field of the jet results in non-thermal continuum emission through synchrotron emission. This results in polarized emission, as seen at radio to optical wavelengths, which serves as a direct indicator of the state of the magnetic field. Emission at high energies (hard \xray{} to \gray{}) is attributed to Inverse Compton (IC) scattering between the synchrotron particle population and ambient photon fields (e.g. the synchrotron photons themselves, external photons from the accretion disk, broad line region or circum-nuclear dust, \cite{Bottcher2012}). The underlying blazar emission processes are described by the Spectral Energy Distribution (SED), which is studied through simultaneous multi-wavelength observations, while the study of simultaneous optical polarization and multi-wavelength data makes it possible to relate changes in the magnetic field properties of the jet, the driver of the observed emission, to changes in the SED. These results can then be compared with blazar emission models, e.g. \cite{Katarzynski2008, HESS2009, Fermi2010, HESS2012, Chandra2015}.
 
Here, we investigate the long-term optical polarization of the archetypal blazar PKS~2155--304 in order to gain insight into the relation between the magnetic field structure and the spectral energy distribution of the source. 

\section{Observations and Analysis}
\begin{figure}[ht!]
  \centering
  \includegraphics[trim = 2.7cm 6.8cm 2.cm 7.4cm, clip ,scale=.7]{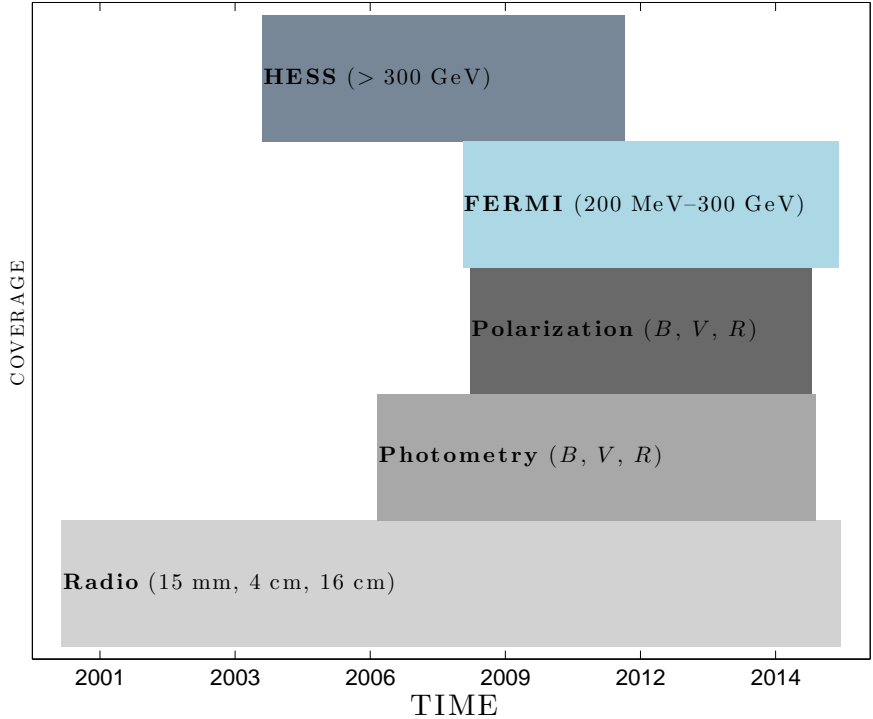}
   \caption{\small{Graphical representation of the spectral band coverage for the available observations.}}
  \label{fig:mwlcoverage}
\end{figure}

\begin{figure}[ht!]
  \centering
  \includegraphics[trim = 2.8cm 1.7cm 2.65cm 1.9cm, clip ,width=1\textwidth]{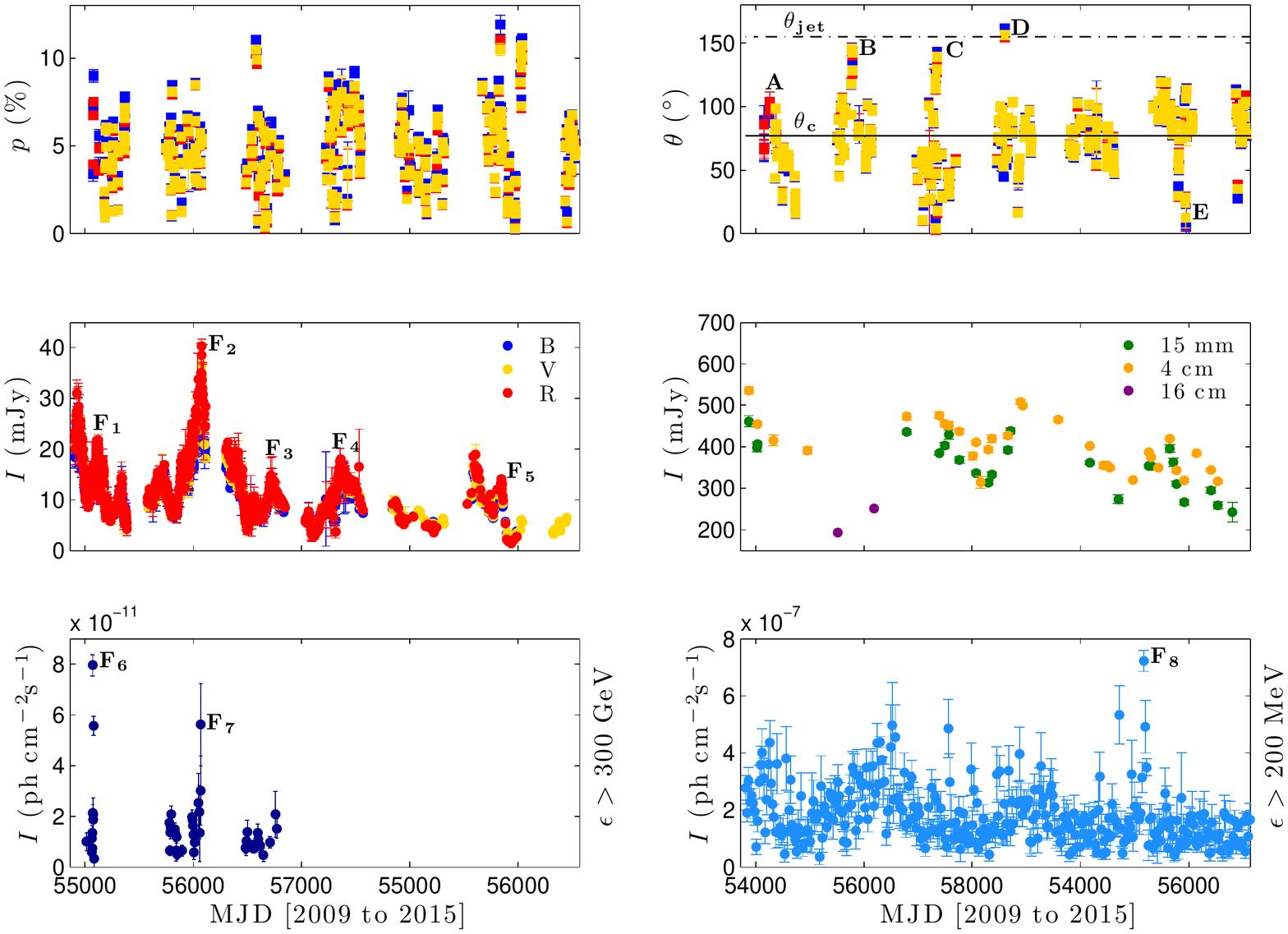}
   \caption{\small{The multi-wavelength variability of \pks{} between July 2009 and December 2014. The top panel illustrates the behaviour of the polarization degree ($p$) on the left and the polarization angle ($\theta$), on the right. The position angle of the resolved radio jet, $\theta_\mathrm{jet}$, and the persistent optical polarization component, $\theta_\mathrm{c}$, is superposed. The optical and radio intensities ($I$) are displayed on the left and right, respectively, of the middle panel. The spectral band of the observations are listed in the legend, which are the \Bf, \Vf and \Rf --band, and at 15 mm, 4 cm and 16 cm. The units of measurement are indicated in brackets. The gamma-ray intensities are displayed in the bottom panel for photon energies $\epsilon>300$ GeV (left) and $\epsilon> 200$ MeV (right). All measurements represent daily averages, except for the HE emission, which represents weekly averages (due to the low sensitivity of the instrument).}}
  \label{fig:obs}
\end{figure}

The polarization of electromagnetic radiation is observable as the polarization degree ($p$) and the polarization angle ($\theta$). The polarization degree measures the fraction of the total intensity that is polarized, while the polarization angle gives the orientation of the electric field vector\footnote{In the plane of the sky, $\theta$ is measured from the North Celestial Pole towards the Celestial East.}. Optical spectro-polarimetric measurements of \pks{} between July 2009 and December 2014 were obtained from long-term monitoring of the source at the Steward Observatory (Arizona, USA)\footnote{The results from Steward monitoring campaign is publicly available at \url{http://james.as.arizona.edu/~psmith/Fermi/}.}. Overlapping photometric observations were obtained from three optical monitoring programmes: the Small and Moderate Aperture Research Telescope System\footnote{The SMARTS archive is publicly available at \url{http://www.astro.yale.edu/smarts/glast/home.php}.} (SMARTS \cite{Bonning2012}), Automatic Telescope for Optical Monitoring (ATOM \cite{HESS2014}) and Robotic Optical Transient Search Experiment (ROTSE \cite{Kastendieck2011}). High-energy (HE) gamma-ray observations (photons with energies from 200 MeV -- 300 GeV) were obtained with the space-based Fermi gamma-ray telescope\footnote{The Fermi database is publicly available at \url{http://fermi.gsfc.nasa.gov/ssc/}.}, while Very High-Energy (VHE, photons with energies exceeding 300 GeV) observations were obtained from the long-term monitoring campaign of the ground-based \hess{} telescope \cite{HESS2014}. Total integrated radio intensity measurements were obtained at 15 mm, 4 cm and 16 cm from the Australia Telescope Compact Array (ATCA) Calibrator Survey. Figure ~\ref{fig:mwlcoverage} illustrates the overlap in spectral band coverage available for \pks{} between 2009 and 2014. The multi-wavelength light curves of the source during this period are displayed in Fig.~\ref{fig:obs}. 

Significant changes in the strength of the polarized emission is detected, from nearly unpolarized emission with a polarization degree of $p\sim 0.5\%$ to a maximum of $p\sim 13\%$. Variations in the polarization are typically seen to occur on timescales of a few days, similar to the high-energy emission, whereas the polarization angle, total optical and radio intensity appear to fluctuate on timescales of a few months. Note that we are unable to resolve short-timescale variations for the high-energy \gray{} emission since the HE light curve has a time-resolution of 7 days. However, some structure is detected on timescales of a few months.

From Fig.~\ref{fig:obs}, it can be seen that the multi-wavelength time series of \pks{} exhibit a number of interesting features, including: large changes in the polarization angle (indicated by A, B, C, D and E), a number of minor optical flares (symbolized by $F_\mathrm{1}$, $F_\mathrm{3}$, $F_\mathrm{4}$ and $F_\mathrm{5}$), as well as a prominent optical flare ($F_\mathrm{2}$) and strong to moderate \gray{} flares ($F_\mathrm{6}$, $F_\mathrm{7}$ and $F_\mathrm{8}$). The outbursts are associated with periods during which the polarization angle undergoes large rotations, typically $\gae 80^\circ$. For example, during A, the polarization angle rotates by 88\dgr{} and the source experiences simultaneous optical ($F_\mathrm{1}$) and VHE ($F_\mathrm{6}$) flares, while the polarization degree experiences a minimum of $\sim 1\%$. Similar correlated behaviour is observed for B, C, D and E. A very prominent optical flare ($F_\mathrm{2}$), which is accompanied by the VHE \gray{} flare $F_\mathrm{7}$, is seen during B, the optical flares $F_\mathrm{3}$ and $F_\mathrm{4}$ are observed during C and D, respectively and the HE \gray{} flare seen during E appears to be accompanied by optical flaring activity at $F_\mathrm{5}$. These results provide evidence supporting a correlation between large polarization angle rotations and the appearance of optical and, sometimes, \gray{} flares. 

Finally, if the jet has a persistent polarized component then the average values of the Stokes parameters, which completely describe the polarization, will be displaced from the origin of the Stokes plane. For \pks, this yields a polarization degree $p_\mathrm{c}=3.7\pm 0.2\%$ and a polarization angle $\theta_\mathrm{c}=77^\circ\pm2^\circ$. Since the polarization angle measures the orientation of the electric field vector, the orientation of the corresponding magnetic field (which is orthogonal to the electric field) is $\theta_\mathrm{c}+90^\circ=167^\circ$. This means that the magnetic field of the persistent jet component effectively lies parallel to the direction of the resolved radio jet, for which the position angle is $\theta_\mathrm{jet}\sim 160^\circ$ \cite{Piner2008, Piner2010}. It is also noteworthy that the most prominent optical flare ($F_\mathrm{2}$) occurs directly after the observed polarization angle aligns with the direction of the jet. 

\section{Concluding Remarks and Future Plans}
We have collected optical polarization measurements for \pks{} spanning a period of roughly five years, during which time the source experienced several optical and gamma-ray flares. A comparison of the evolution of the polarization angle during these outbursts suggest that some of the flaring activity may be related to the orientation of the polarization angle relative to the position angle of the jet. Future work will involve a more detailed analysis of a specific flaring event, wherein we will track the evolution of the polarization angle and spectral energy distribution before, during and after the flare. These changes in the SED will then be compared to the orientation of the polarization angle, which will allow us to investigate how the physical processes that produce the observed radiation is related to the magnetic field of the emission region. The results of this study (to be published in 2017) can then be compared to blazar emission models that also make predictions for the polarization, e.g. \cite{Zhang2014, Zhang2015}.

\end{document}